# Highly Polarized Light Emission from 6T@BNNT Nanohybrids

A. Badon[1], C. Allard[2], F. Fossard[3], A. Loiseau[3], L. Cognet[1], E. Flahaut[4], N. Izard[5], R. Martel[6], and E. Gaufrès[1*]

[1] Laboratoire Photonique Numérique et Nanosciences, Institut d'Optique, CNRS UMR5298, Université de Bordeaux, F-33400 Talence, France

[2] Département de génie physique, Polytechnique Montréal, Montréal, Québec H3C 3A7, Canada

[3] Laboratoire d'Étude des Microstructures, ONERA-CNRS, UMR104, Université Paris-Saclay, BP 72, 92322 Châtillon Cedex, France

[4] CIRIMAT, Université de Toulouse, CNRS, INPT, UPS, UMR CNRS-UPS-INP N°5085, Université Toulouse 3 Paul Sabatier, Bât. CIRIMAT, 118, route de Narbonne, 31062 Toulouse cedex 9, France

[5] Laboratoire Charles Coulomb, UMR5221 CNRS-Université de Montpellier, 34095 Montpellier, France

[6] Département de chimie, Université de Montréal, Montréal, Québec H3C 3J7, Canada

*Correspondence: etienne.gaufres@u-bordeaux.fr

**Abstract:**

**The polarized fluorescence emission of organic fluorophores has been extensively studied in photonics and is increasingly exploited in single molecule scale bio-imaging. Expanding the polarization properties of compact molecular assemblies is, however, extremely challenging due to depolarization and quenching effects associated with the self-aggregation of molecules into the sub-nanometer scale. Here we demonstrate that Boron Nitride Nanotubes (BNNTs) can act as a 1D host-template for the alignment of encapsulated α-sexithiophene (6T) inside BNNTs, leading to an optically active 6T@BNNT nanohybrid. We show that the fluorescence from the nanohybrid is strongly polarized with extinction ratios as high as 700 at room temperature. A statistical analysis of the 6T orientation inside BNNTs with inner diameter up to 1.5 nm shows that at least 80% of the encapsulated 6Ts exhibit a maximum deviation angle of less than 10° with respect to the BNNT axis. Despite a competition between molecule-molecule and molecule-BNNT adsorption in larger BNNTs, our results also show that more than 80%**

**of the molecules display a preferential orientation along the BNNT axis with a deviation angle below 45°.**

**Main:**

Polarized light-matter interaction in molecules offers additional functionalities of high interest in various applications, such as bio-imaging,[1] solar energy harvesting[2,3] and quantum photonics[4], to name just a few. In particular, common organic dyes, such as rod-like polythiophenes, combine huge photon absorption/emission transitions between well-oriented transition dipoles with bright polarized emission at the single molecule level.[5–9] Yet, the unique polarization properties and bright emission of a single molecule is hard to translate to larger assemblies of that molecule because the self-assembly process generally prevents extended alignment of the molecular dipoles and hence lowers its emissivity. The formation of highly polarized dye aggregates requires: (1) Specific molecular ordering and alignment in the assembly and (2) a fine control of the interaction between molecules (e.g. van der Waals, hydrogen bonding, etc.). One emblematic molecular architecture for polarized emission is the head-to-tail stacking of dyes. This so-called J-aggregate favours intermolecular coupling that leads to bright lower emission states, giving higher fluorescence quantum yields, narrower red-shifted emission bands and a strong polarization dependency.[6–8] Targeting this specific molecular arrangement is extremely challenging because the self-organization (or stacking) of common molecules rarely results in the desirable head-to-tail ordering. Many self-aggregation schemes have been investigated using supramolecular assemblies with or without selective bonding and many examples have been found thanks to specific adsorption mechanisms at surfaces or in solutions.[9–11] Other approaches based on a host/guest concept and using nanoporous crystals, such as zeolites and more recently metal-organic frameworks (MOF), have also attracted significant interest for templating molecular dipoles.[12,13] While these approaches have greatly expanded our understanding of the polarization effects related to the supramolecular structure of the emitters, the synthesis and manipulation of large assemblies of molecules for highly polarized emission remains a limiting step for wide-spread applications. Since their discovery, single-walled carbon nanotubes (SWCNTs) have been used as a host of molecules and as template for their assembly [18,19] and there are many examples related to the encapsulation of organic dyes (Dyes@CNT).[18,20,21] Thanks to the 1D crystalline architecture, the nanotube offers a hollow centre that is readily available for molecular assembly and alignment. Their cavity is characterized by a diameter and chiral angle distributions with an impressive structural regularity along the whole nanotube axis. Although the SWCNT is structurally an ideal template for J-aggregates, this research has produced conflictual results on the emission properties of the dyes due to the presence of an efficient energy transfer processes between the photo-excited dye and the SWCNT host ($E_g < 1$ eV). It was recently shown that the fluorescence from most of the dyes are efficiently quenches by the SWCNT.[22,23] However, our group has recently addressed the quenching issue by replacing SWCNTs with boron nitride nanotubes (BNNTs).[24] Similar to a SWCNT, the BNNT has hollow cavity in the centre and presents a 1D crystalline structure having an inner diameter distribution ranging between 1-2 nm. However, they differ significantly by a strong dielectric behaviour up to the bandgap energy close to 6 eV.[25–27] Our work showed that the BNNT sheet protects the luminescence properties organic dyes with clear signatures of bright J-like aggregation states and enhanced photostability.

Here we explore the fluorescence polarization properties of α-sexithiophene (6T) molecules encapsulated inside BNNTs (6T@BNNTs). By using aberration corrected High Resolution Transmission Electronic Microscopy (HRTEM) and fluorescence imaging of individual 6T@BNNTs, we observe that the molecules are remarkably well aligned along the BNNT axis for inner diameter below 3 nm. We report that this 1D ordering is independent from the length of the BNNT and of the filling level. We finally show that a complex adsorption and stacking competition induces molecular disorder in the larger inner diameter BNNTs, which complicates the polarization dependency when the inner diameter is above ~3 nm.

**Results and discussion.**

The sample preparation begins by opening raw BNNTs (BNNT LLC supplier) using mechanical grinding and ultrasound treatments, followed by a thermal purification at 800°C in air to remove the content of elemental boron and other h-BN sub-products. Purified BNNTs are then dispersed in DMF by ultrasounds and a small fraction of the solution is either spin-coated on a Si/SiO$_2$ substrate (previously patterned with localisation marks) or drop casted onto a Molybdenum TEM grid coated with a holey SiO$_2$ membrane. The BNNTs are then liquid-filled with 6Ts by placing the sample in a solution of 6T dissolved in toluene at 115°C and at a concentration of 5x10$^{-6}$ M. The last step of this liquid phase encapsulation consists in removing the non-encapsulated 6T molecules using toluene rinsing cycles at room temperature. Residues of 6Ts that are not encapsulated are finally removed using an oxidative piranha solution (or a mild oxygen plasma in some experiments). More details about the sample preparation can be found in the supplementary information file.

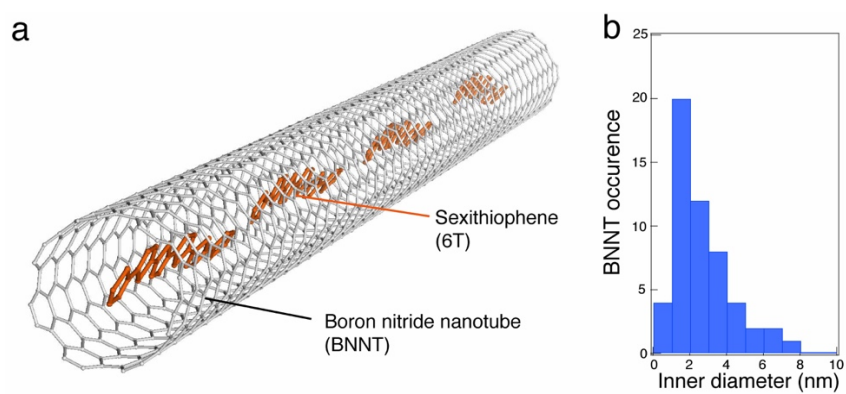

**Figure 1**: **a** Schematic illustration of α-sexithiophene (6T) molecules aligned inside a boron nitride nanotube (BNNT). **b** Diameter distribution of BNNTs probed on the grid using Transmission Electron Microscopy. Note that BNNTs generally have 2 to 4 walls; wall numbers superior to 5 and diameters larger than 10 nm are seldom found in our samples.

Figure 1a presents a schematic representation of 6T molecules aligned inside a BNNT. In practice, the measurements of the inner diameter, $d_{inner}$, of our BNNT samples is obtained statistically using aberration corrected High Resolution Transmission Electron Microscopy (ac-HRTEM) experiments and the results indicate a relatively broad distribution, ranging from 0.7 nm to 7 nm with an occurrence maximum centred at about 2 nm (see Fig 1b). Considering the 1 nm length of an elongated 6T molecule, this large distribution of the size of the BNNT cavity should induce various stacking configurations that can compete with the head-to-tail alignment required for highly polarized emission. To investigate this configuration distribution, we performed polarized fluorescence imaging at the single nanotube level of 6T@BNNTs lying on Si/SiO$_2$ substrate (Fig. S1). Briefly, a polarized 532 nm laser (1:1000 polarization ratio) was collimated onto the sample using the 60x objective to homogeneously photo-excite individual 6T@BNNTs within a ~ 45x45 µm$^2$ field-of-view. A combination of half and quarter wave plates allowed to excite the sample with a well-defined polarization state (linear or circular with controlled orientations). The emitted fluorescence was collected and analysed through a polarization beam splitter prism cube acting as the "analyser" and a 660±5 nm band-pass filter. Details on the optical set up are available in the supplementary information.

Figure 2a shows a representative series of polarized fluorescence images obtained from individual 6T@BNNTs recorded using different excitation polarization conditions (blue arrows). The top series displays the fluorescence of randomly oriented 6T@BNNTs having lengths comparable to the optical resolution of our apparatus (~1 µm) and excited with circularly polarized light at a laser wavelength of 532 nm. Note that the circular polarization condition ensures that all orientations are simultaneously detected with our setup. The middle and bottom series present the response of two orthogonal 6T@BNNTs having relatively longer lengths of ~4 µm. The first one presents a homogeneous fluorescence intensity along the BNNT axis while the second displays spotted emission due to a partial filling of 6T dyes. The two 6T@BNNT are excited with a linearly polarized light oriented approximatively along the 6T@BNNT axis.

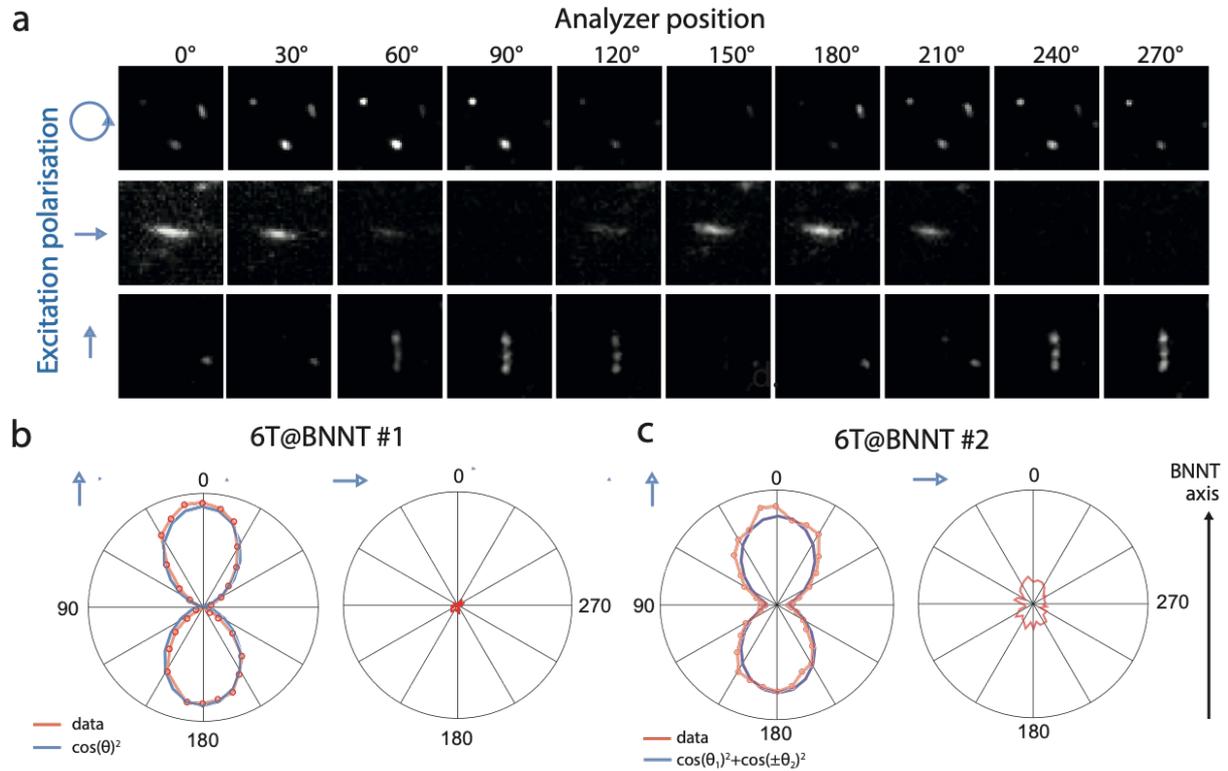

**Figure 2: a** Polarized fluorescence images of short (< 1 μm) and long (~ 4 μm) individual 6T@BNNT candidates deposited on Si/SiO$_2$ substrate. The polarization angle of the signal ranges from 0° to 270° with an incremental step of 30° between each image. The blue arrows indicate the direction of the excitation polarization for the series: circular, horizontal and vertical polarization from top, middle to bottom, respectively. **b-c** Polar plots of the fluorescence intensity for two 6T@BNNT as a function of the analyser position and for two excitation polarizations, i.e. along and orthogonal to the nanotube axis. The excitation wavelength is 532 nm. The fitting curves are represented by a blue line.

For all of these different cases, the periodic and severe extinction of the signal at specific angles unambiguously demonstrates a highly polarized fluorescence emission from all of these 6T@BNNT candidates. These examples feature representative cases of what is typically observed in our samples (see other examples in the supplementary information file). It's interesting to note that strong polarization effects is systematically observed in our samples and that they occur independently of 6T@BNNT length, diameter and filling rate.

To better analyse the observed polarization dependencies, we plotted in Figure 2b-c the integrated intensity of two typical 6T@BNNTs as a function of the analyser polarization angle, θ. Exciting a given candidate with a linearly polarized light approximatively parallel to the nanotube axis maximizes the signal to noise (s/n) ratios and this condition was selected for these plots. The first 6T@BNNT (Figure 2b, left) displays a highly polarized emission with an ultimate extinction ratio of 700, which is only limited by the s/n ratio. In cross-polarized excitation conditions (Figure 2b right), the signal is also near zero. The large extinction of the signal and the good fit with a $\cos^2\theta$ function (blue line) is consistent with a strong and mono-disperse alignment of the molecules along the nanotube axis. This result highlights the ability of BNNTs to impose an homogeneous molecular alignment along the BNNT axis, even over many thousands of molecules.

While perfect alignment is noted in most 6T@BNNT, we found cases such as in Figure 2c where the polarization dependency appears more complex. While the plot denotes an overall strong polarization of the emission along the BNNT axis, the shape of the polar lobes is slightly altered compared to the behaviour in Figure 2b. In those cases, an incomplete extinction of the signal is found at 90° and the experimental fit function clearly requires more than a simple $\cos^2\theta$ function (Figure 2c blue line). These results strongly suggest that, albeit limited, a distribution of misaligned molecular sub-population takes place inside those specific BNNT individuals. We note that the polarization dependency seems to be characterized by preferential orientation of the polarized intensity at higher angles.

For a better understanding of the polarization patterns in 6T@BNNT fluorescence, we imaged many individual 6T@BNNTs supported on a grid using an aberration corrected HRTEM operated at 80kV. Figure 3a presents some representative HRTEM images of 6T@BNNT nanohybrids having few walls and diameters ranging from 0.8 nm to 3.5 nm. As already reported before, [24] an overall ordering of the 6T molecules is clearly observed for BNNTs with the smallest inner diameters. While a modelling of these images is not yet available, one can readily see in these HRTEM images that a subset of the molecules adopts larger angles relative to the nanotube axis and this is more preeminent as inner diameters increases. Some examples are highlighted by white arrows. To clarify this behaviour, we quantitatively measured the deviation angle for 757 molecules confined inside a statistical assembly of 50 different BNNTs having inner diameters ranging from 0.8 to 4.5 nm.

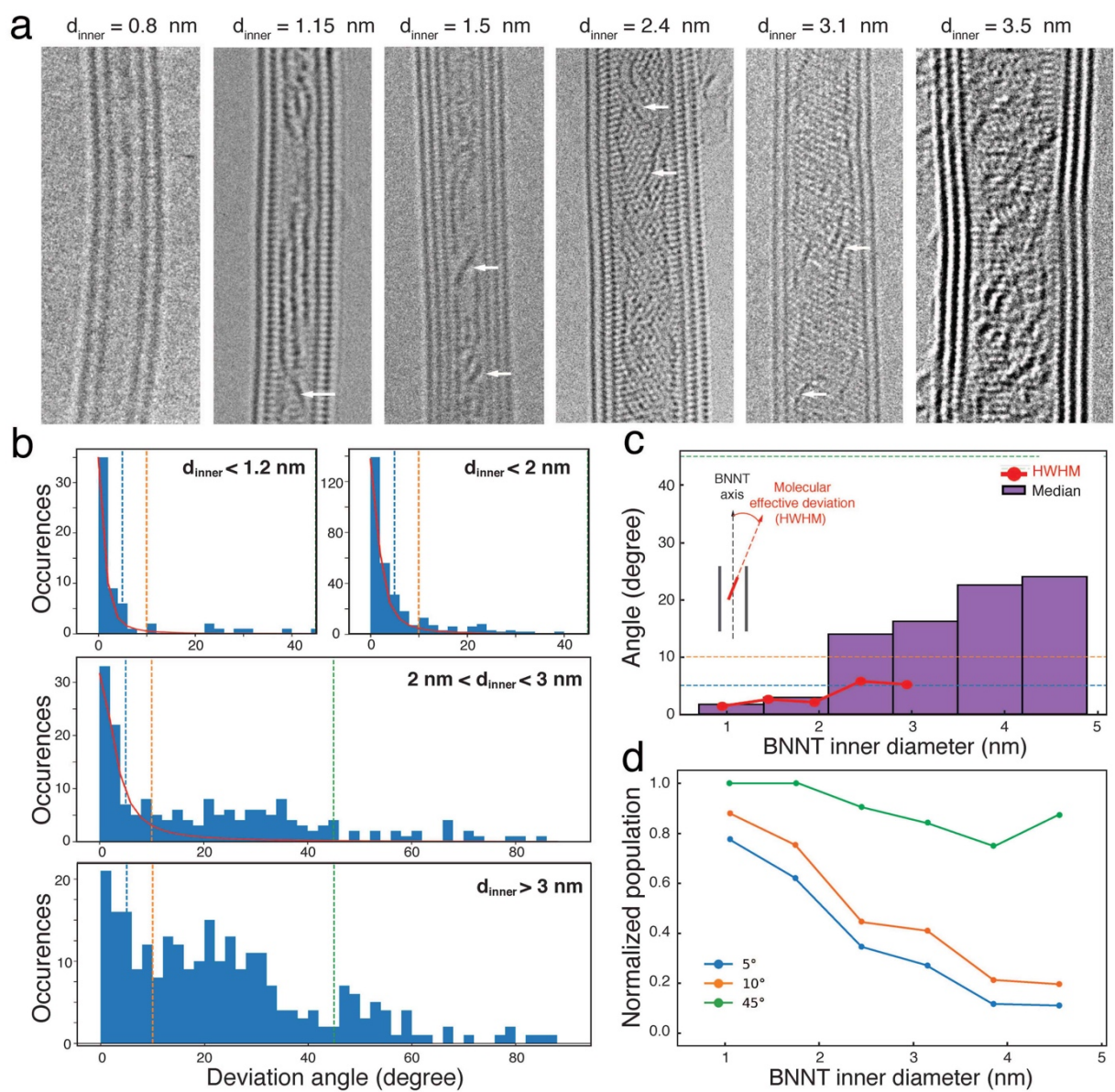

**Figure 3. a** HRTEM images recorded at 80kV of 6T@BNNTs with inner diameters, $d_{inner}$, between 0.8 and 4.5 nm. The white arrows highlight some molecules with a deviation angle in the range of 20°- 40°. **b** Distributions of the deviation angles (absolute values) between the molecules and the BNNT axis for different inner diameter subgroups. **c** Red line: Evolution as a function of the inner diameter of the Half Width Half Maximum (HWHM) of the deviation angle distributions extracted from a Lorentzian fit of the distribution of the values angles in b Purple histograms: Median value of the angle distributions as a function of the inner diameter. **d** Normalized evolution of the 6T populations having a deviation angle of less than 5°, 10° and 45° as a function of the inner diameter.

Figure 3b presents the occurrences of deviation angles obtain by HRTEM for $d_{inner}$ < 1.2 nm, $d_{inner}$ < 2 nm, 2 nm < $d_{inner}$ < 3 nm and $d_{inner}$ > 3 nm, respectively. Two different distributions appear clearly in the statistics: the first one has a sharp distribution close to 0°, which is representative of the smallest inner diameters. The second distribution is very broad and ranges from 10° to 60° and their prevalence is for $d_{inner}$ > 2 nm. The mean deviation angle of the latter distribution is roughly 20°- 40°.

While the observation of a preferential angle distribution in small (compared to the molecule length) inner diameter BNNTs is not surprising, the results reveals a cross over between perfectly aligned assemblies to mostly aligned distribution characterized by a sub population of molecules having 20-40° orientations relative to the nanotube axis. This evolution is clearly highlighted in the Half Width Half Maximum (HWHM) plot in Figure 3c, extracted from Lorentzian fits applied to the deviation angle distributions for each 0.5 nm diameter interval (see supporting information file for details). An interesting feature is that the HWHM remains below 7 ° up to diameters of 3 nm, a distance that is more than twice the length of a 6T molecule. For $d_{inner}$ > 3nm, a simple Lorentzian function cannot take in account the second distribution at 20°- 40° (see Figure 3b middle) and the limited statistics in this diameter range does not allow the use of a second Lorentzian function. To overcome this limitation, we also plotted in Figure 3c the median value of the distributions and observed that this value never exceeds 25° for 6T@BNNT with $d_{inner}$ < 5 nm.

The last panel of Figure 3 show the normalized populations of the 6T molecules orientations, characterized by a maximum angle deviation of 5°, 10° and 45°, respectively. Two important trends in the distribution are identified in this plot. For BNNTs with $d_{inner}$ < 1.5 nm, we see that more than 80% of the molecules have deviation angles less than 10°. For larger BNNTs, the distribution shows that 80% of the 6T molecules display a deviation angles below 45°.

Although the orientation distributions become more complex at $d_{inner}$ > 2 nm, the results are consistent with the polarization emission patterns observed in Figure 2, which show two main signatures: a population that is strictly polarized along the nanotube axis and a second population exhibiting a more complex polarization dependency with a preferential distribution aligned in the direction of the nanotube axis. The ability of the BNNT wall to adsorb molecules seems to force a cross-over in the orientation distribution for $d_{inner}$ above 3 nm. This is particularly noticeable on the two HRTEM images in Figure 3a of the 6T@BNNTs exhibiting $d_{inner}$ = 2.4 nm and 3.1 nm. Hence, the templating mechanism inferred by interaction with the nanotube sidewalls seems to compete with intermolecular interactions. A first hypothesis to explain the cross over is that the molecules tend to form their natural tilted stacking like (also called herringbone) such as in aggregated polymorphic 6T crystals.[28–30] Indeed, the weak interactions between 6T enable various tilt angles of 6T molecules ranging from 45° in the low temperature phase to 65° in the high temperature phase. Another driving force to tilt the orientation is the effect of the chiral angle of the BNNT itself, which may force the 6T stacking in tilted orientation relative to the nanotube axis. Preferential chirality angles of 15°-30° are reported in large multi-walled BNNTs,[31] but a flat distribution from 0° to 30° is rather common in single- and double-walled BNNTs.[32]

In summary, we have measured the polarization distribution of individual nanohybrids made of α-sexithiophene encapsulated inside boron-nitride nanotubes having inner diameter ranging from 0.8 nm to 4.5 nm. The results underline a correlation between the preferential alignment of the molecules observed with transmission electron microscopy and the polarization dependency of the luminescence emission relative to the nanotube axis. The remarkable ability of BNNTs for creating 1D aggregates of perfectly aligned 6T molecules is highlighted by a complete extinction of the emission intensity at 90° and in a cross polarization configuration. The overall alignment of the molecules and BNNT axis enables highly polarized patterns of fluorescence emission from all observed 6T@BNNTs. Nevertheless, the samples reveal a cross over between a perfectly aligned distribution to a mixed distribution having larger angles when the inner diameter is in the range of 2-3 nm. Complementary to the application interest of the 6T@BNNTs for polarized photonics, a fine analysis of the polarization properties of 6T inside BNNTs can be used as a powerful indirect probe of the inner diameter of BNNTs.

**Aknowledgments**

E.G. acknowledges funding from the Marie-Skłodowska-Curie-IF 706476-BrightPhoton and from CNRS starting package funding, the GDRi Graphene and Co, the GDRi multifunctional nano for travel support. This work received financial support from the Natural Sciences and Engineering Research Council of Canada (NSERC) under grants RGPIN-2019-06545 and RGPAS-2019-00050 and Canada Research Chairs (CRC) programs. L.C. acknowledges financial support from the Agence Nationale de la Recherche (ANR-15-CE16-0004-03) and ITMO Cancer AVIESAN within the framework of the Cancer Plan (18CPl21-00). The authors warmly acknowledge G. Wang for support on the Cs-corrected TEM of MPQ - Paris Diderot University and METSA support for access to the Cs-corrected TEM of MPQ. A.B. acknowledges financial support from the Agence Nationale de la Recherche (ANR-17-CE18-0026)

**Data Availability**

The data that support the findings of this study are available from the corresponding author upon reasonable request.

**Competing Interest Statement**

The authors declare no competing financial interests.

Supporting information file for:

# Highly Polarized Light Emission from 6T@BNNT Nanohybrids

A. Badon[1], C. Allard[2], F. Fossard[3], A. Loiseau[3], L. Cognet[1], E. Flahaut[4], N. Izard[5], R. Martel[6], and E. Gaufrès[1*]

[1] Laboratoire Photonique Numérique et Nanosciences, Institut d'Optique, CNRS UMR5298, Université de Bordeaux, F-33400 Talence, France

[2] Département de génie physique, Polytechnique Montréal, Montréal, Québec H3C 3A7, Canada

[3] Laboratoire d'Étude des Microstructures, ONERA-CNRS, UMR104, Université Paris-Saclay, BP 72, 92322 Châtillon Cedex, France

[4] CIRIMAT, Université de Toulouse, CNRS, INPT, UPS, UMR CNRS-UPS-INP N°5085, Université Toulouse 3 Paul Sabatier, Bât. CIRIMAT, 118, route de Narbonne, 31062 Toulouse cedex 9, France

[5] Laboratoire Charles Coulomb, UMR5221 CNRS-Université de Montpellier, 34095 Montpellier, France

[6] Département de chimie, Université de Montréal, Montréal, Québec H3C 3J7, Canada

*Correspondence: etienne.gaufres@u-bordeaux.fr

### 1- Materials

Boron nitride nanotubes (BNNT) were provided by BNNT LLC. Only reagent grade solvents were used. 3-aminopropyltriethoxysilane (APTES) (99%) and α-sexithiophene (6T) were purchased from Sigma-Aldrich and used as received.

### 2- Sample preparation

**Preparation of surfaces with localisation markers**

Si/SiO$_2$ surfaces (100 nm oxide) were dehydrated in a YES oven. OIR 674 resist was spin-coated on the substrate at 4000 rpm for 30s and then baked for 60s at 90°C. Photolithography was done with an exposition dose of 36.5 mJ/cm$^2$ using a Karl Suss MA6 mask aligner. After development, 5 nm of titanium and 20 nm of gold were deposited on the substrate using e-beam deposition. Lift-off was done in warm acetone.

**Preparation of surfaces with BNNTs**

BNNTs were provided from BNNT LLC and purified as previously described [1]. Briefly, BNNTs were treated in nitric acid and/or at high temperature in air, and then centrifuged at 12 350g. The purified BNNTs were dispersed in DMF at a concentration of 50 μg/mL.

Si/SiO$_2$ surfaces with localisation markers were sonicated for 10 min each in acetone and isopropanol and immersed in a piranha solution (3:1) for 5 min. Substrates where rinsed copiously with miliQ water and dried with N$_2$. Then, an APTES treatment was applied to the surfaces using a vapor-phase method. The substrates were placed on glass slides suspended above a crystallization dish with 1 mL of APTES, and the dessicator was vacuum pumped for one minute and sealed for 30s. The APTES layer was annealed in air for 30-45 minutes at about 100 ºC in a conventional oven. The BNNT solution was then spin-coated onto the surface at 3000 rpm for 30s.

**Encapsulation of 6T on surfaces**

The Si/SiO$_2$ surface with BNNTs was annealed in a vacuum oven at 800°C for 1h and immediately used for encapsulation. The encapsulation was done in a flask equipped with a condenser, under reflux. The concentration of the 6T solution was fixed at $5 \times 10^{-6}$ M and the encapsulation was carried out for 24h at 115°C. After refluxing, the surface was rinsed with fresh solvent and IPA and dried with N$_2$. Figure S1 shows an AFM image of the prepared surface after encapsulation.

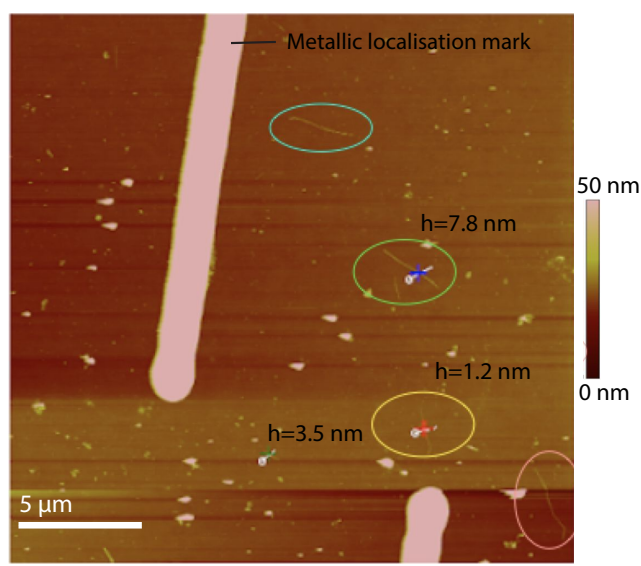

Figure S1 - AFM image of 6T@BNNTs on a substrate with metallic localisation marks

**Encapsulation of 6T in BNNT for TEM grid (Mo/SiO$_2$)**

About 20uL of BNNT solution in DMF was drop-casted on a Molybdenum grid with SiO$_2$ membrane decorated with holes. Before encapsulation, the TEM grid was annealed under vacuum at 800°C for two hours. The grid was inserted in the 6T encapsulation solution at 115°C for 6 hours. Following the encapsulation, the grid was rinsed for a few seconds in DMF, cleaned using an oxygen plasma (100W, 10 minutes) and a piranha treatment (2 minutes) to completely remove the excess of non-encapsulated dyes.

## 3- Optical set up for polarized fluorescence experiments

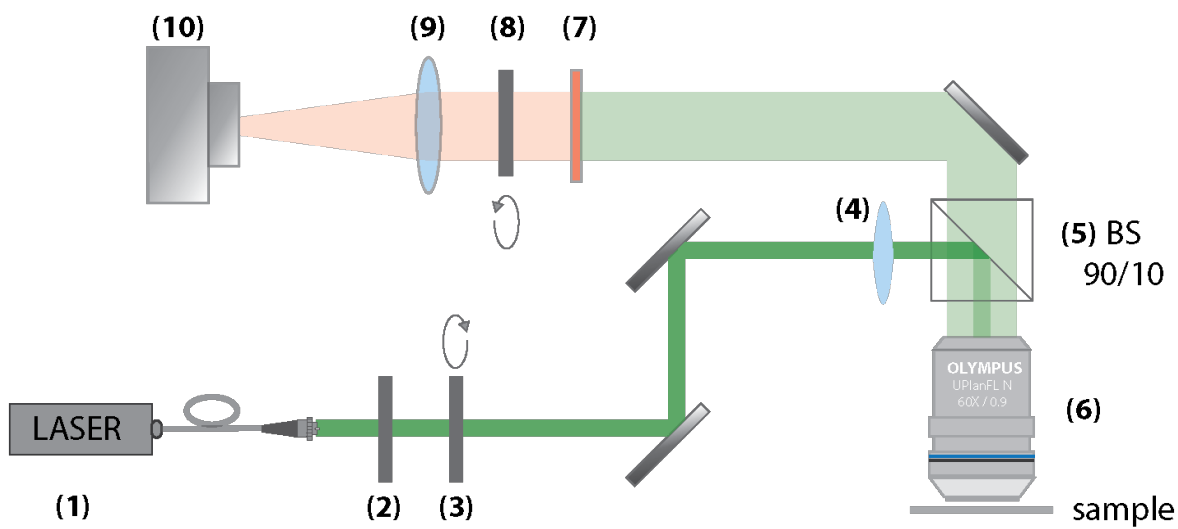

**Fig S2. List of the elements used for polarized micro-fluorescence experiment.**
(1) Oxxius Laser: 532 nm, 100mW, polarization extinction ratio (1:1000)
(2) Glan prism
(3) $\lambda/2$ or $\lambda/4$ wave plates
(4) Non polarized prism cube beam-splitter (Tr90:R10)
(5) Focalisation lens (300 mm)
(6) Objective 60X, NA 0.9
(7) Band-Pass filter 660 nm ± 5 nm
(8) Polarized prism cube as "analyser"
(9) Focalisation lens (75mm)
(10) Nitrogen cooled Pylon camera (Princeton) (1340x400)

## 4- Fit of deviation angle distribution

The angle distribution was adjusted using a non-linear least-square minimization and curve-fitting package [2], using the built-in Lorentzian model. The center of the Lorentzian was fixed to an angle of 0 degree, while all the other fitting parameters were free. The Half-Width Half-Maximum (HWHM) was directly extracted from the fitting parameters.

[1] Allard, C., Schué, L., Fossard, F., Recher, G., Nascimento, R., Flahaut, E., ... & Gaufrès, E. (2020). Confinement of Dyes inside Boron Nitride Nanotubes: Photostable and Shifted Fluorescence down to the Near Infrared. Advanced Materials, 2001429.
[2] http://dx.doi.org/10.5281/zenodo.11813

## 5- Ac-HRTEM experiment

The high-resolution images presented in figure 3 were performed at 80 kV on a JEOL ARM microscope, equipped with an aberration corrector.